\documentclass[pre,twocolumn,showpacs,preprintnumbers,amsmath,epsfig,amssymb,aps]{revtex4}

\usepackage{amsmath}
\usepackage{amssymb}
\usepackage{graphicx}
\usepackage[usenames]{color}
\usepackage{xspace}

\newcommand{\dd}[2][t]{\frac{d#2}{d#1}\xspace}

\newcommand{\coeff}[2]{\ensuremath{#1_\text{\tiny{#2}}}}
\newcommand{\betaS}{\coeff{\beta}{S}}
\newcommand{\betaR}{\coeff{\beta}{R}}
\newcommand{\muON}{\coeff{\mu}{ON}}
\newcommand{\muOFF}{\coeff{\mu}{OFF}}
\newcommand{\alphaP}{\coeff{\alpha}{P}}
\newcommand{\alphaS}{\coeff{\alpha}{S}}
\newcommand{\alphaR}{\coeff{\alpha}{R}}
\newcommand{\alphaN}{\coeff{\alpha}{N}}
\newcommand{\nS}{\coeff{n}{S}}
\newcommand{\nR}{\coeff{n}{R}}
\newcommand{\kS}{\coeff{k}{S}}
\newcommand{\kR}{\coeff{k}{R}}

\newcommand*{\tofig}{Fig.}
\newcommand*{\totab}{Table}
\newcommand*{\toeq}[1][{}]{Eq#1.}

\graphicspath{{mygraphics/}}

\begin{document}

\title{Messenger RNA Fluctuations and Regulatory RNAs Shape the
Dynamics of Negative Feedback Loop}

\vspace{.5cm}

\author{Mar\'{\i}a \surname{Rodr\'{\i}guez Mart\'{\i}nez}}
\altaffiliation[Present address: ]{%
  Center for Computational Biology and Bioinformatics, Columbia University,
  New York, USA.}
\affiliation{Department of Physics of Complex Systems,
  Weizmann Institute of Science, Rehovot, Israel.}
\affiliation{Department of Molecular Genetics,
  Weizmann Institute of Science, Rehovot, Israel.}
\author{Jordi Soriano}
\affiliation{Departament d'ECM, Facultat de F\'{\i}sica,
       Universitat de Barcelona, Barcelona, Spain.}
\author{Tsvi Tlusty}
\email[Corresponding author: ]{tsvi.tlusty@weizmann.ac.il.}
\affiliation{Department of Physics of Complex Systems,
  Weizmann Institute of Science, Rehovot, Israel.}
\author{Yitzhak Pilpel}
\email[Corresponding author: ]{pilpel@weizmann.ac.il}
\author{Itay Furman}
\affiliation{Department of Molecular Genetics,
  Weizmann Institute of Science, Rehovot, Israel.}

\begin{abstract}

Single cell experiments of simple regulatory networks can markedly
differ from cell population experiments. Such differences arise from
stochastic events in individual cells that are averaged out in cell
populations. For instance, while individual cells may show sustained
oscillations in the concentrations of some proteins, such
oscillations may appear damped in the population average. In this
paper we investigate the role of RNA stochastic fluctuations as a
leading force to produce a sustained excitatory behavior at the
single cell level. Opposed to some previous models, we build a fully
stochastic model of a negative feedback loop that explicitly takes
into account the RNA stochastic dynamics. We find that messenger RNA
random fluctuations can be amplified during translation and produce
sustained pulses of protein expression. Motivated by the recent
appreciation of the importance of non--coding regulatory RNAs in
post--transcription regulation, we also consider the possibility
that a regulatory RNA transcript could bind to the messenger RNA and
repress translation. Our findings show that the regulatory
transcript helps reduce gene expression variability both at the
single cell level and at the cell population level.
\end{abstract}



\maketitle

\section{Introduction}

The growing interest in biological noise has led to many efforts to
measure gene expression at the single cell level \cite{Elowitz2002,
BarEven2006, Newman2006}, revealing a very distinct dynamics when
compared to population cell experiments \cite{Raj2009}.  In two
well--studied examples, the p53--mdm2 regulatory network and the
NF--$\kappa$B signaling pathway, sustained oscillations are observed
in single cells following activation signals \cite{LevBarOr2000,
Tiana2002, Nelson2004}, while cell population experiments only show
damped oscillations \cite{Lahav2004, GevaZatorsky2006,
Batchelor2008}. In both cases the core circuit consists of a
negative feedback loop, one of the most common network designs,
where the active transcription factor promotes the transcription of
its own repressor.

Stable oscillations are not trivially generated in a single negative
feedback loop \cite{Griffith1968}. A loop composed of only two
agents does not oscillate for plausible macroscopic equations.
Sustained oscillations require at least three agents, where the
third one introduces a time delay that repeatedly causes the system
to overshoot or undershoot above and below the steady state
\cite{Elowitz2000}. Some models achieve sustained oscillations by
introducing \textit{ad hoc} time delays to reproduce those that a
system incurs when manufacturing the various molecular components
\cite{GevaZatorsky2006, Ma2005}. The dynamics can also be enriched
by considering combinations of negative and positive feedback loops
\cite{Vilar02, Ciliberto2005, Zhang2007}, bistable switches
\cite{Francois2005}, or by inheriting oscillatory signals from
upstream regulators \cite{Ma2005, Batchelor2008}.

In this paper we show that the stochastic fluctuations in gene
expression in a negative feedback loop can produce sustained
pulses of protein expression. It has been suggested that
protein fluctuations are driven by underlying messenger RNA (mRNA)
fluctuations \cite{BarEven2006, Newman2006, Thattai2001}. We show
that the mRNA stochastic fluctuations can be amplified during
translation and induce a sustained excitatory behavior characterized
by a series of sustained anti--correlated pulses in the expression
of the positive and negative regulator of the loop.

Noise induced oscillations have already been found in other systems.
Oscillations in a circadian clock consisting of a combination of a
positive and a negative feedback loop are enhanced by the intrinsic
biochemical noise \cite{Vilar02}. Resonant amplification of the
stochastic fluctuations can lead to cycling behavior in the Volterra
system \cite{Mckane2005} and in self--regulatory genes
\cite{Mckane2007}.
Here we show that a simple negative feedback loop consisting of an
activator protein and its repressor is capable of producing protein
pulses when the stochastic fluctuations of the mRNA are taken into
account. This result does not rely on having a large number of
molecules or on the particular statistical properties of the noise,
neither depends on upstream pulsating signals or couplings to
additional loops.

Recently several experimental studies have shown that gene
expression occurs in bursts of transcriptional activity
\cite{Golding2005, Chubb2006, Raj2006}. These bursts are usually
ascribed to random upstream events, such as chromatin remodeling or
random promoter transitions. Here we demonstrate that sustained
pulses of protein expression can be produced merely by the
stochastic nature of mRNA kinetics.

In view of the crucial significance of mRNA fluctuations, we asked
ourselves how gene expression can be accurately regulated in the
noisy cellular environment.
Regulation of gene expression is a complex, multi--layered process
that involves many different players. Since their discovery more
than a decade ago, regulatory RNAs (termed ``regRNAs'' in this
paper) have emerged as key regulators in virtually all the cellular
processes studied to date. regRNAs are non--coding RNAs that
regulate gene expression by base--pairing to a partially or fully
complementary mRNA target. MicroRNAs (miRNA) \cite{Bartel2004} and
antisense RNA \cite{Lapidot2006} are two examples of regulatory
RNAs.

In the mammalian genomes regulatory RNAs often share the same
transcriptional regulation as their targets, giving rise to a
diversity of feed--forward loops \cite{Cawley2004,Shalgi2007}. Such
pairs of target and regulatory RNA transcripts are found to be
co--regulated, co--expressed or inversely expressed more frequently
than expected by chance, presumably due to sharing of common
transcription factors
\cite{Lapidot2006,Shalgi2007,Chen2005,Katayama2005}. In particular,
some transcription factors have been found to bind to overlapping
transcript pairs, thus potentially coupling the regulation of a
coding gene and its regulatory RNA \cite{Cawley2004}. It was
suggested that one potential purpose of such design is to filter
transcription noise \cite{Lapidot2006}.

In this work we consider the case in which the positive regulator in
the loop transcribes both mRNA and regRNA.
The latter could be either miRNA or an antisense.
Additionally, we assume that the regulatory transcript
binds to the mRNA and prevents its translation, but without
promoting its degradation. We show that the presence of regRNA reduces the
excitability of the system by increasing its capacity to buffer the noise.

\section{Models}

\begin{figure}[!ht]
\begin{center}
\includegraphics[width=8cm]{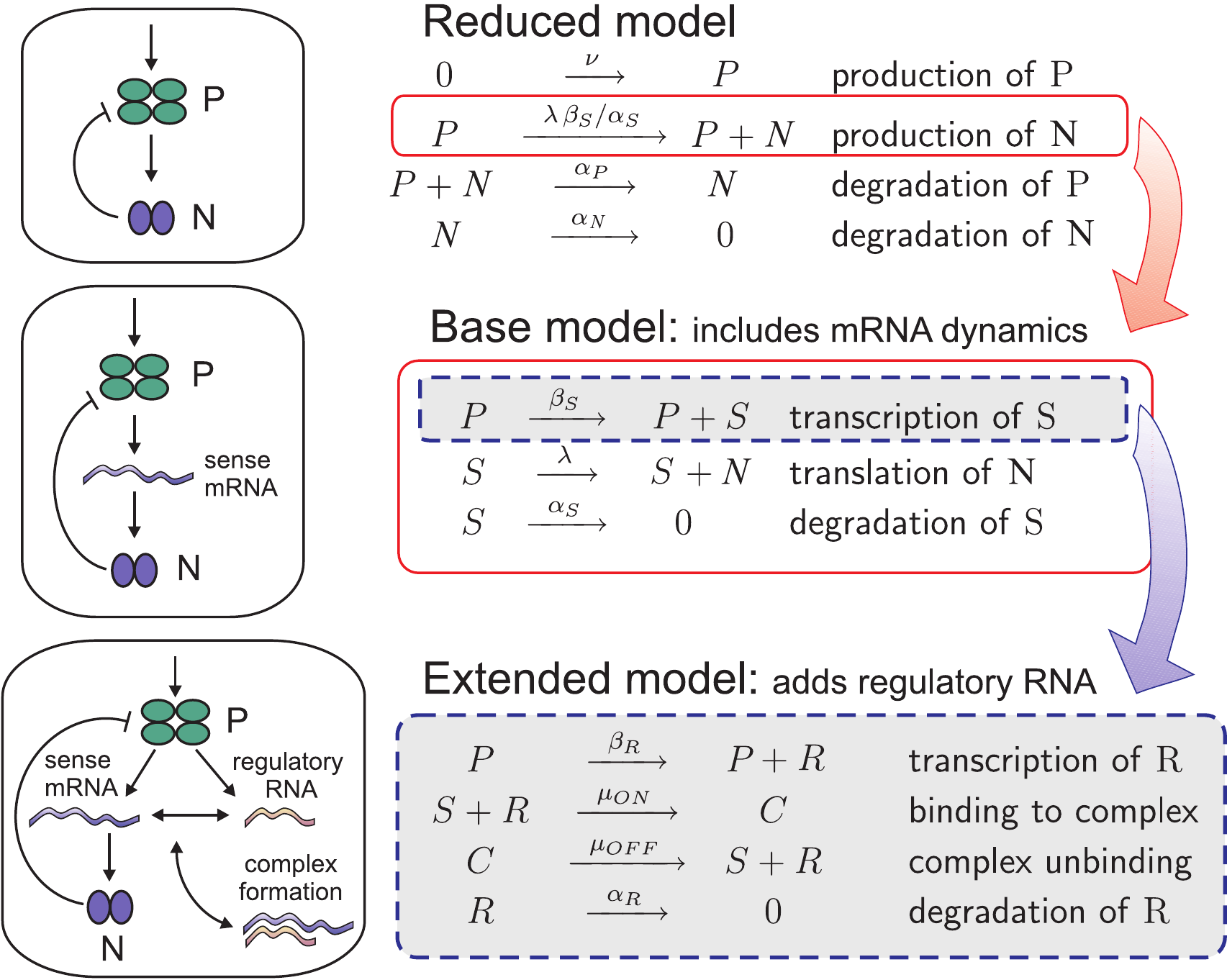}
\caption{\label{fig:models} (Color online) Schematic representation
of the reduced, base, and extended models together with the chemical
reactions. The reduced model, making up the simplest feedback loop,
considers only the production and degradation of the proteins ($P$)
and ($N$), and assumes that mRNA is in quasi--equilibrium. The base
model includes mRNA dynamics by expanding the chemical reaction of
the production of $N$ (red box).  The additional chemical reactions
describe the transcription of sense mRNA ($S$), the translation of
$N$, and the degradation of $S$. Finally, the extended model puts
into play regulatory RNAs, $R$ (dashed blue box), by adding their
transcription, degradation, and the formation and degradation of a
sense--regRNA complex $C$. }
\end{center}
\end{figure}

We consider three alternative designs of a negative feedback loop.
The main loop is composed of a transcription factor (the positive
regulator, $P$) and its repressor (the negative regulator, $N$).  We
assume that in response to some external cellular signal, $P$ has
been activated and is promoting the transcription of $N$. Two of the
designs also model the transcript dynamics. In the three models, $P$
is degraded at the post--translational level via protein--protein
interaction with $N$. Schematic representations of the three models,
as well as the individual chemical reactions, are shown in
\tofig~\ref{fig:models}.
The description of the biochemical parameters and their range of
variation are summarized in \totab~\ref{tab:parameters}.
\begin{table}
  \caption{\label{tab:parameters}%
    Symbols used as parameters of the models,
    their values or range of variation, and their description.}
  \begin{ruledtabular}
    \begin{tabular}{crl}
        symbol & \multicolumn{1}{c}{value} & \multicolumn{1}{c}{meaning} \\
         & \multicolumn{1}{c}{(min.$^{-1}$)} & \\
         \colrule
        $\nu$     & $200$             & Transcription factor ($P$) induction\\
        $\betaS$  & $0.05$--$10$      & mRNA ($S$) induction                \\
        $\betaR$  & $0.01$--$20$      & anti-sense ($R$) induction          \\
        $\lambda$ & $0.05$--$50$      & protein ($N$) translation           \\
        $\muON$   & $0.001$--$10$     & sense--regRNA binding                \\
        $\muOFF$  & $0.001$--$10$     & sense--regRNA unbinding              \\
        $\alphaP$ & $0.10$            & $N$-assisted decay of $P$           \\
        $\alphaS$ & $0.03$            & sense auto--degradation              \\
        $\alphaR$ & $0.03$            & regRNA auto--degradation             \\
        $\alphaN$ & $0.10$            & protein ($N$) auto-degradation      \\
        $\nS$ & $3$\footnotemark[1]   & Hill's coeff., sense transcription  \\
        $\nR$ & $1$\footnotemark[1]   & Hill's coeff., regRNA transcription \\
        $\kS$ & $500$\footnotemark[1] & Threshold, sense transcription      \\
        $\kR$ & $300$\footnotemark[1] & Threshold, regRNA transcription     \\
    \end{tabular}
     \end{ruledtabular}
\footnotetext[1]{dimensionless parameters}
\end{table}
The values adopted correspond to typical mammalian cells.

\subsection{The base model}

The base model consists of the two proteins regulators $P$ and $N$,
and an mRNA transcript $S$.  (The latter will be interchangeably
refered to as ``transcript'', ``sense transcript'', or ``sense
mRNA''.)
$P$ promotes the transcription of $S$,
which in turn encodes for the negative regulator $N$.
The deterministic equations are:
\begin{subequations}
  \label{eq_net_I}
    \begin{align}
        \dd{P} &= \nu - \alphaP N P,
           \label{eq_net_I:1} \\
        \dd{S} &= \betaS \frac{P^{\nS}}{P^{\nS} + \kS^{\nS}} -\alphaS S,
           \label{eq_net_I:2} \\
        \dd{N} &= \lambda S - \alphaN N.
           \label{eq_net_I:3}
    \end{align}
\end{subequations}
We assume that the cooperative binding of $n_s$ molecules of $P$ at
the promoter site of $N$ are required for efficient transcription,
and we model the transcription rate with a Hill's function. The
translation rate of $N$ is proportional to the transcript level,
$S$. Finally, the rate of the $N$--mediated degradation of $P$ is
proportional to the concentration of both proteins.

\subsection{The extended model: Incorporation of regulatory RNA}

This model introduces a regulatory RNA (regRNA), denoted $R$, that
targets $S$. $P$ promotes the transcription of both $S$ and $R$
transcripts. $R$ regulates $S$ by base--pairing to it, thus creating
a hybrid $S$-$R$ complex, $C$. We assume that the complex molecule
can unbind but not degrade, \textit{i.e.}, it returns to the system
both complementary RNA transcripts.

The new set of equations describing the model are given by:
\begin{subequations}
    \label{eq_net_III}
    \begin{align}
        \dd{P} &=  \nu - \alphaP N P,    \label{eq_net_III:1}\\
        \dd{S} &= \betaS\frac{P^{\nS}}{P^{\nS} + \kS^{\nS}}
                  - \muON R S + \muOFF C - \alphaS S,
                                         \label{eq_net_III:2} \\
        \dd{R} &= \betaR\frac{P^{\nR}}{P^{\nR} + \kR^{\nR}}
                  - \muON R S + \muOFF C - \alphaR R,
                                         \label{eq_net_III:3}\\
        \dd{C} &= \muON R S - \muOFF C,  \label{eq_net_III:4}\\
        \dd{N} &= \lambda S - \alphaN N. \label{eq_net_III:5}
    \end{align}
\end{subequations}
The extended model contains $14$ parameters described in
\totab~\ref{tab:parameters}. In this work we focus on analysing how
the temporal behavior of the protein levels is influenced by the
mRNA and regRNA transcription rates, the translation rate, and the
$S$--$R$ complex formation and destruction.  Therefore, we
systematically explored how the variation in $\betaS$, $\betaR$,
$\lambda$, $\muON$, and $\muOFF$, affect the emergent system
bahvior.  We maintained the remaining parameters constant throughout
the simulations; random exploration showed that changes in their
value did not affect the behavior of the system appreciably.

\subsection{The reduced model: Neglected RNA kinetics}

Finally, to isolate the effect of RNA fluctuations we consider a
reduced model where the RNA transcripts are assumed to be in
quasi--equilibrium. From a molecular perspective, this model
corresponds to the limit where the time scales associated with RNA
transcription and degradation are much shorter than those associated
with protein production and degradation.  Thus, the transcript
levels adjust rapidly to changes in $P$ and $N$, and $\frac{dS}{dt}
\approx 0$ in \toeq~(\ref{eq_net_I:2}). The equations become:
\begin{subequations}
    \label{eq_net_II}
    \begin{align}
        \dd{P} &= \nu - \alphaP N P, \label{eq_net_II:1} \\
        \dd{N} &= \frac{\lambda\betaS}{\alphaS}
                  \frac{P^{\nS}}{P^{\nS} + \kS^{\nS}} - \alpha_N N.
                  \label{eq_net_II:2}
    \end{align}
\end{subequations}

\section{Numerical algorithm and data analysis}

We modeled the stochastic behavior at the single--cell level using
the Gillespie algorithm \cite{Gillespie1977}. To quantify the
importance of the stochasticity we solved the deterministic
equations and compared their trajectories with the stochastic
dynamics for each set of chemical parameters in all three models.
The deterministic equations always reached a steady state, which is
the same in all three models for a given set of parameters. We chose
this value as initial condition of the stochastic simulations in
order to minimize the initial transient period. Each single
stochastic simulation represents a possible cell realization, the
equivalent of a single--cell experiment in this description. To
simulate the behavior of a random distribution of cells, we computed
$300$ realizations with different random seeds for each set of
chemical parameters. The typical length of each simulation was
$2000$ min.

For each molecular species $X$, where $X$ stands for
$P,N,S,R,\text{and}, C$, we computed the average molecular count
$\bar{X}=\langle X(t) \rangle_t$ (averaged across all simulations
and over time), and the coefficient of variation, $\mathcal{V}$,
defined as the ratio between the standard deviation and the average
level
\begin{equation}
  \mathcal{V}(X) = \frac{\langle [X(t)-\bar{X}]^2\rangle_t^{1/2}}{\bar{X}}.
  \label{eq:CoV}
\end{equation}
$\mathcal{V}$ significantly increases with the amplitude of pulses
of $X$, therefore it provides a general signature of the
stochasticity and strength of the pulsations for the three models.

To identify the pulsating dynamics in more detail, we computed the
normalized autocorrelation function \cite{Loinger2007b}, given by
\begin{equation}
  C(\tau) = \frac{\langle
    \left[ X(t)-\bar{X}\right] \left[X(t+\tau)-\bar{X}\right] \rangle_t}{%
    \langle\left[X(t)-\bar{X}\right]^2\rangle_t}.
\end{equation}
The autocorrelation function for a each model and parameters was
averaged over $300$ realizations of the initial conditions.

Additionally, we compared the results of the autocorrelation with the
power spectrum, defined as $S(k)$ = $\langle F(k)F(-k)\rangle$, where
$F(k)$ is the Fourier Transform of $X(t)$ and $\langle ...\rangle$
denotes average over realizations.

\section{Results}

\begin{figure*}
\begin{center}
\includegraphics[width=14cm]{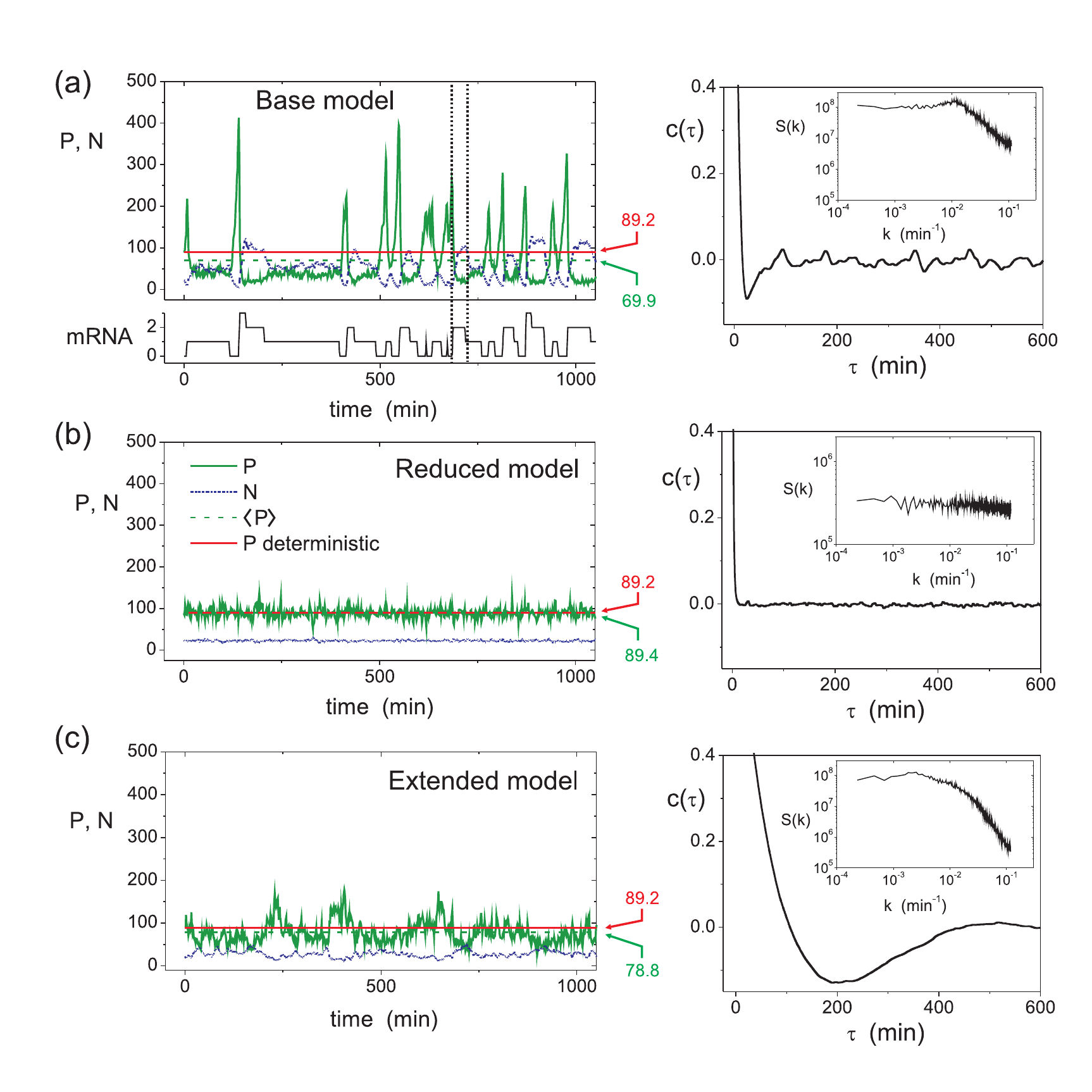}
\caption{\label{fig:examples} (Color online) Examples of stochastic
    simulations in the (a) base, (b) reduced, and (c) extended
    models. Left panels: copy number of $P$ (thick green), $N$ (dotted blue) and
    mRNA (black). The average $P$ copy number (dashed green line) is
    compared to the deterministic trajectory of $P$ (solid red line). For
    clarity, their average values are indicated on the right. The
    vertical dashed lines in (a) illustrate the correlation between
    pulses of $N$ and mRNA. Right panels: detail of the
    corresponding autocorrelation functions $C(\tau)$ and power
    spectra $S(k)$ (inset), averaged over $300$ runs.}
\end{center}
\end{figure*}

\tofig~\ref{fig:examples} provides a snapshot of the stochastic
behavior of the three models.
For each model, an example of a typical simulation is shown on the
left panels, while the corresponding correlation functions and power
spectra (averaged over $300$ runs) are depicted on the right panels.
Overall, the base model is characterized by a clear excitatory
behavior in the positive regulator $P$.  This is confirmed by the
peaked correlation function, which provides a characteristic period
of the pulses.  At the other extreme, the reduced model completely
lacks pulses, and the correlation function rapidly decays to zero.
Finally, the extended model shows a more complex behavior: the time
evolution is characterized by a combination of short pulses and
long--term fluctuations.  The correlation function has a smooth,
oscillating shape, that corresponds to the long--term fluctuations,
and a characteristic period of the pulses cannot be identified.

The detailed description of the results for each model, as well as an
analysis of the conditions for the presence or lack of excitations, are
provided next.

\subsection{Stochastic fluctuations of RNAs can produce pulses of
protein expression}

We first considered the base network, \toeq[s]~(\ref{eq_net_I}),
which contains the positive regulator $P$, the mRNA $S$, and the
negative regulator $N$. \tofig~\ref{fig:examples}(a) shows the
outcome of a typical stochastic simulation.  For comparison, the $P$
deterministic value is also shown. The stochastic and deterministic
simulations exhibited remarkable differences. While the
deterministic concentration was locked in a steady state, the
stochastic integration showed a sustained excitatory behavior, where
$P$ and $N$ were expressed in a series of anti--correlated pulses.
The presence of regular pulses of protein expression in the base
model is revealed by the autocorrelation function $C(\tau)$ of $P$,
as shown in the right panel of \tofig~\ref{fig:examples}(a). The
autocorrelation presents a pulsating behavior that provides
quantitative information about the dynamics of the system. The
negative anticorrelation peak indicates the characteristic width of
the pulses, around $25$ min. The first positive peak provides the
characteristic interval between consecutive pulses, around
$100$ min, and that is in agreement with the maximum in the power
spectrum.

\begin{figure}[!ht]
\begin{center}
\includegraphics[width=8.0cm]{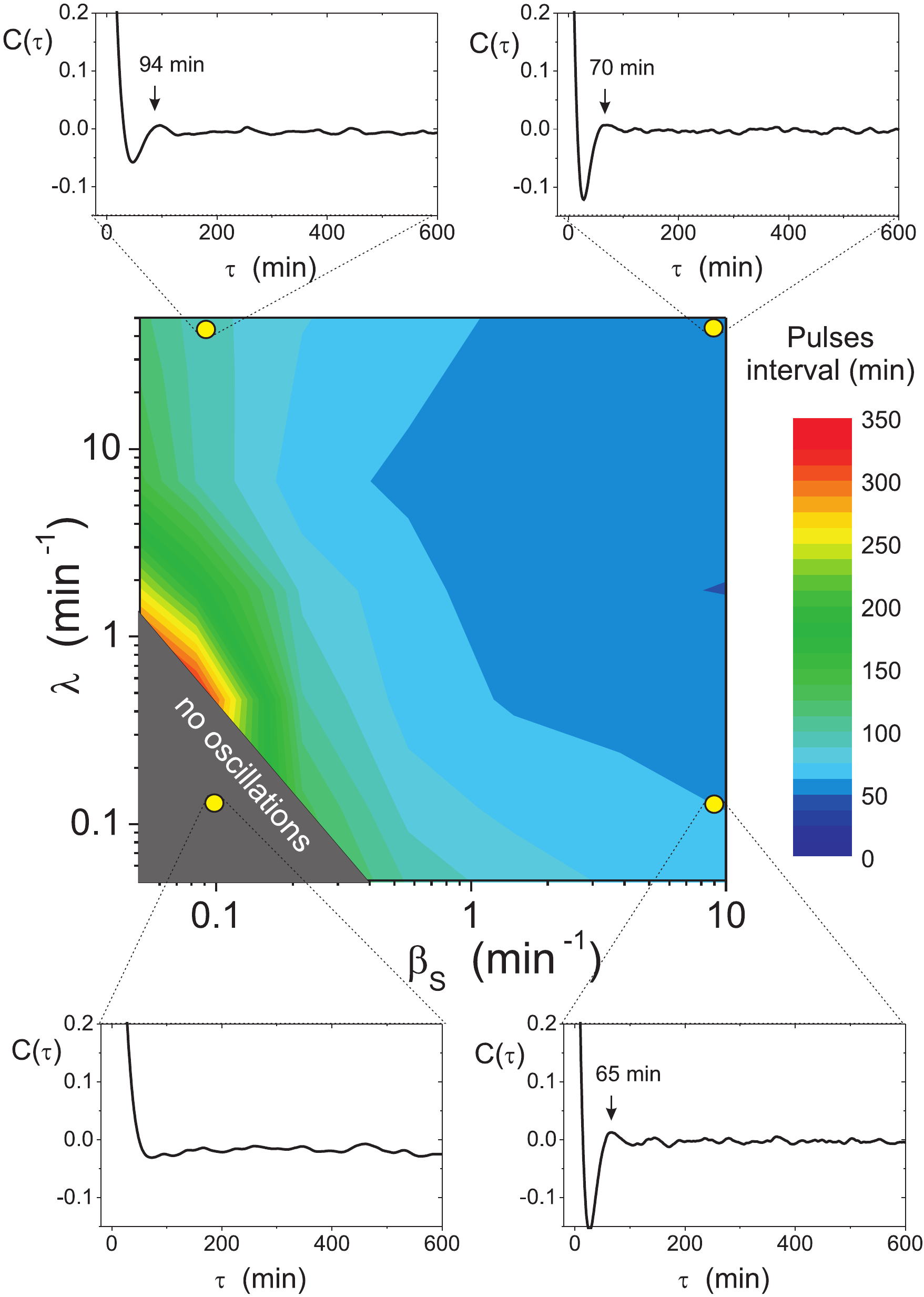}
\caption{\label{fig:timelag} (Color online) Color map of the
characteristic $P$ inter--pulse interval for the base model as a
function of the transcription and translation coefficients ($\betaS$
and $\lambda$ respectively). A sustained excitatory behavior with a
characteristic period appears once minimum translation and
transcription rates are attained. The four surrounding plots, whose
corresponding ($\betaS$,$\lambda$) values are indicated with dots on
the color map, illustrate the behavior of the autocorrelation
function $C(\tau)$, averaged over $300$ runs, for $4$ extreme
combinations of transcription and translation rates.  Except for
very low transcription and translation rates, the autocorrelation
$C(\tau)$ shows peaks of correlation that identify the
characteristic inter--pulse interval.}
\end{center}
\end{figure}

Minimal transcription and translation rates of the negative regulator
$N$ were required to identify an excitatory behavior with an
observable pattern, as illustrated in \tofig~\ref{fig:timelag}.
Below a minimal values of $\nu$ and $\lambda$ the amount of $N$ was
insufficient to implement an effective negative feedback: the pulses
of protein expression became irregular, while the average interval
between consecutive pulses increased.  At extremely low
transcription and translation rates the pulses disappeared
altogether. Above these extreme conditions we observed that in
general a characteristic inter--pulse period emerges in the base
model for all tested sets of parameters (\tofig~\ref{fig:timelag}).
The autocorrelation shows clear peaks that provide a characteristic
period. The average period depends on the transcription and
translation rates, and ranges between $30$ and $200$ min for
realistic rates.

The origin of the pulses can be understood by observing the mRNA
time trace.  As shown in \tofig~\ref{fig:examples}(a), the mRNA is
transcribed in a series of micro pulses that are subsequently
amplified during translation and inherited by $N$. The strong
correlation between the pulses of mRNA and $N$ is evident in the
figure.

We also observed a strong correlation between the mRNA inter--pulse
time lags and the typical width of the pulses of $P$. As seen in
\tofig~\ref{fig:examples}(a), $P$ only reached significant levels
after a substantial decay in $N$ prior to a new pulse of $N$.
Therefore the typical length of the $P$ pulses depends on the decay
time of $N$, which has been taken to be around $5$ min
\cite{Stommel04}. The fast $N$ degradation also explains the small
width of $P$ pulses compared to the typical time gap between
consecutive pulses.

\begin{figure}[!ht]
\begin{center}
\includegraphics[width=7.6cm]{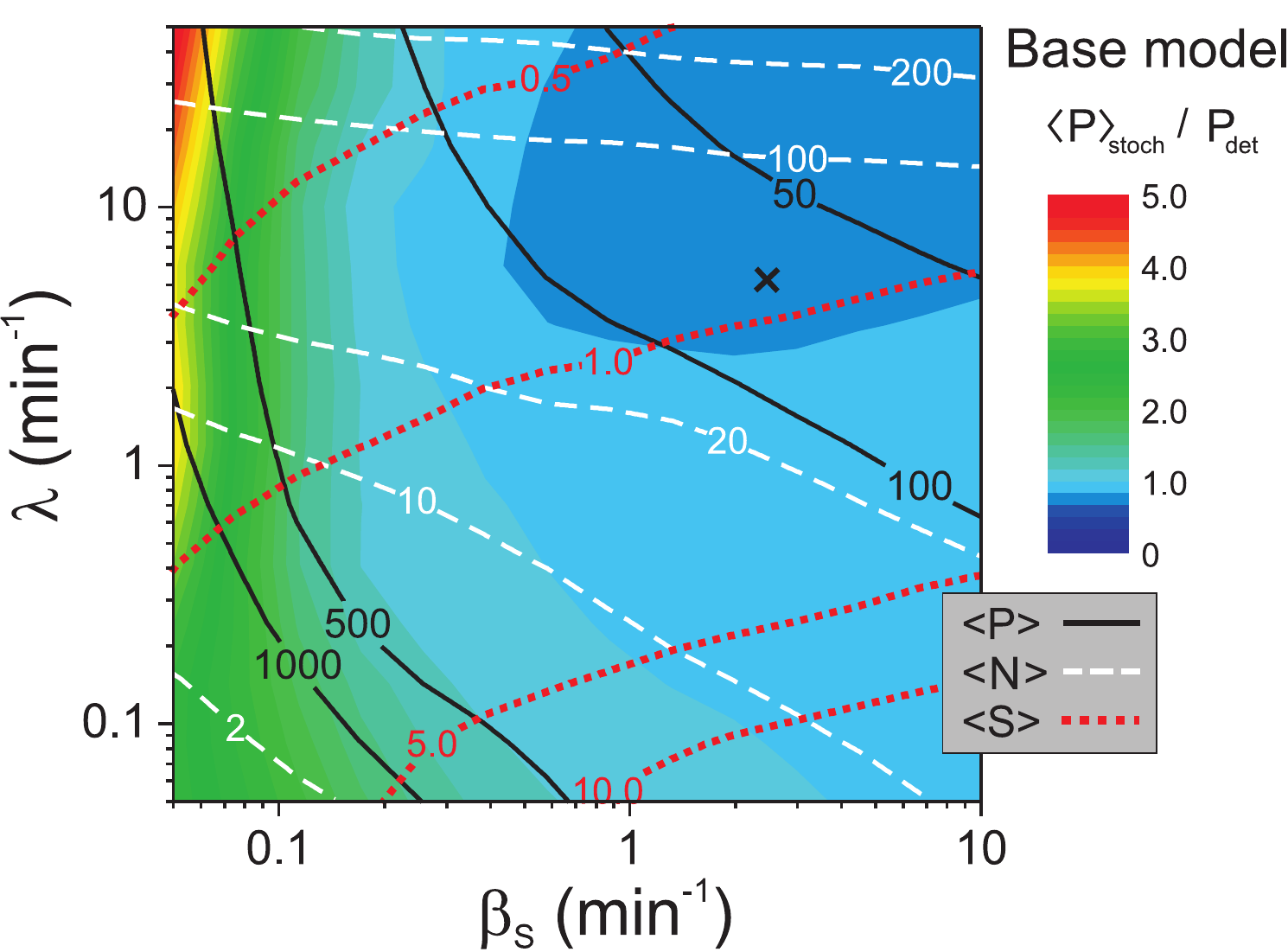}
\caption{\label{fig:ratio} (Color online) Ratio between the $P$
stochastic level averaged over time and over 300 simulations and the
$P$ deterministic steady state in the base model, as a function of
the transcription and translation rates. Contour lines show equal
stochastic levels of $P$ (black), $N$ (dashed white) and mRNA (dotted red). The
black cross indicates the location of the example shown in
\tofig~\ref{fig:examples}(a).}
\end{center}
\end{figure}

\begin{figure}[!ht]
\begin{center}
\includegraphics[width=8cm]{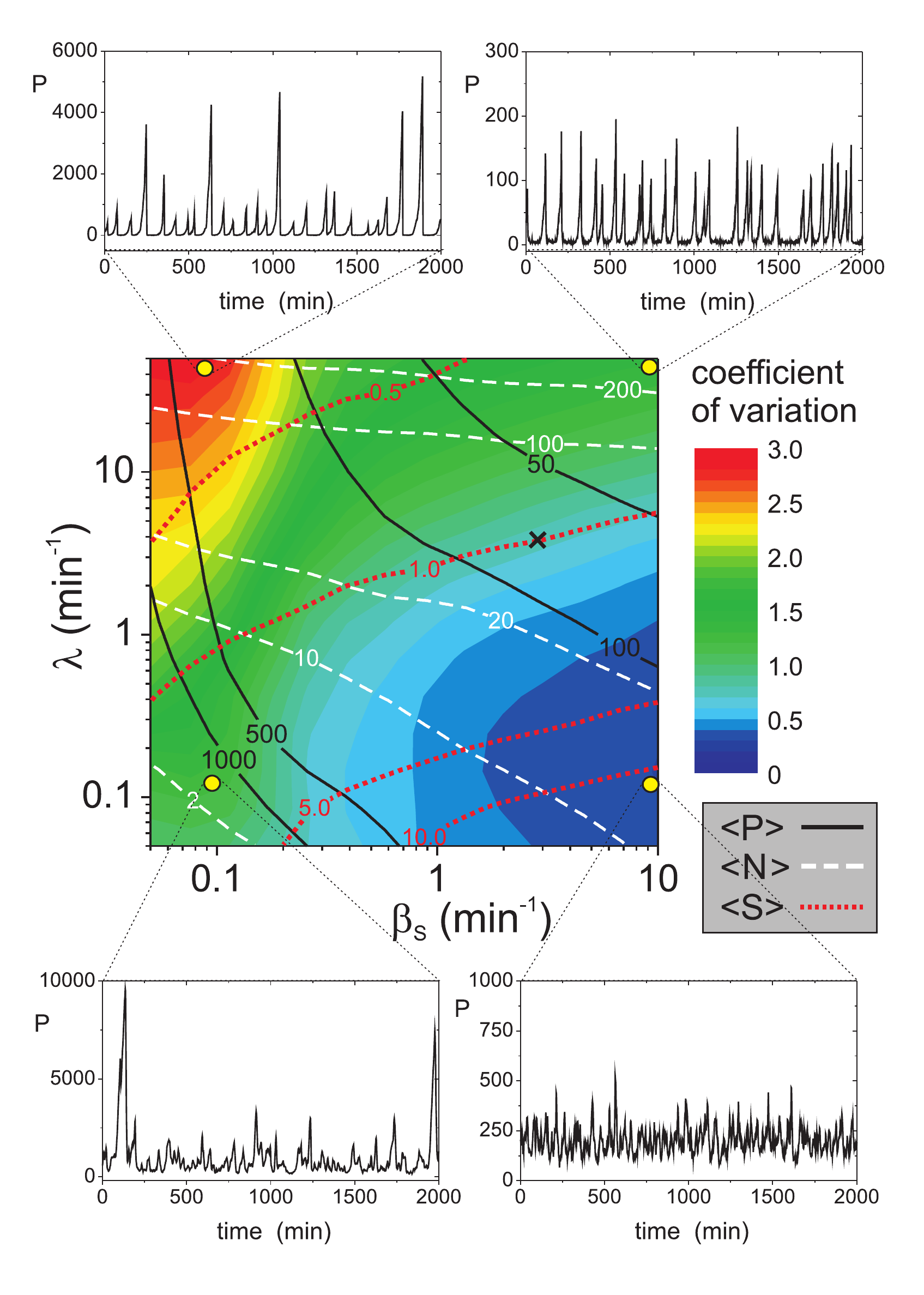}
\caption{\label{fig:CoV} (Color online) Color map of the
coefficient of variation of $P$ in the base model, as a function of
the transcription and translation coefficients ($\betaS$ and
$\lambda$ respectively).  Contour lines show equal average
stochastic levels of $P$ (black), $N$ (dashed white) and mRNA (dotted red). $P$
exhibits maximum coefficient of variation at low $\betaS$ and high
$\lambda$. The black cross indicates the position of the example
shown in \tofig~\ref{fig:examples}(a). The four plots illustrate the
$P(t)$ behavior for four different values of the coefficient of
variation, showing that the pulsating behavior of $P$ is maximum at
high translation rates $\lambda$.}
\end{center}
\end{figure}

Finally, we note that the excitatory behavior of the network crucially
depends on having a low number of mRNAs.  In the base model, the
average number of mRNA copies was below $5$ for most of the explored
parameter space (Figs.~\ref{fig:ratio} and~\ref{fig:CoV}).
This low average value is in agreement with measurements of
transcripts copy number in mammalian cells, where it was found that
many transcripts are present in less than one copy per cell on
average \cite{Carter05}.

\subsubsection{Comparison with the reduced model:
The importance of the mRNA stochastic fluctuations}

To further test the impact of the mRNA stochastic dynamics, we
considered an alternative circuit design, the reduced model
\toeq[s]~(\ref{eq_net_II}), where the mRNA was assumed to be in
quasi--equilibrium and therefore was not explicitly included in the
circuit. \tofig~\ref{fig:examples}(b) shows an example of stochastic
simulation run with the same biochemical parameters as the base
model in \tofig~\ref{fig:examples}(a). The protein dynamics was
considerably different in both models. While the pulses were clearly
present in the base model, no trace of pulses could be found in the
reduced model. Similar conclusions are reached by comparing the
auto--correlation and power spectrum in both models.

Opposed to the base model, intrinsic mRNA fluctuations were not
allowed in the reduced model. Hence, the protein levels showed very
uniform patterns with only small deviations around the deterministic
steady state. The protein levels were also characterized by a very
small coefficient of variation.  The latter is in agreement with the
experimental evidence that when the contribution of external factor
affecting cell--to--cell variability is substracted, noise in
protein expression is dominated by the stochastic production and
destruction of mRNAs \cite{BarEven2006,Newman2006}. We stress that
both networks were simulated with the same algorithm, kinetic
parameters, and initial conditions.

An additional difference between both models was given by the
average of the stochastic simulations. The average, which provides
an insight into cell population dynamics, converged towards the
deterministic trajectory in the reduced model, but not in the base
model. The deviation in the base model became especially important
at low transcription rates.

\tofig~\ref{fig:ratio} shows the ratio of the stochastic $P$ copy
number (averaged over time and over $300$ simulations), and the
deterministic steady state. For transcription rates below $0.1$
transcripts per minute, the deterministic description clearly
underestimated the real circuit behavior. The fact that the
divergence appeared at low transcription rates, and that it was
absent in the reduced model, hints at the mRNA stochastic dynamics
as the source of this divergence. This is another interesting
example of a biological network presenting a deviation between the
stochastic average and the deterministic equations. Such deviations
are commonly expected in non--linear systems and/or systems
containing a small number of molecules, although more generic
systems can also present large deviations \cite{Samoilov2006}.

\subsubsection{Noise is minimized at high transcription and low
translation rates}

To achieve a given protein concentration an organism can adopt diverse
strategies characterized by different transcription and translation
rates.  One strategy consists of producing a few mRNA transcripts and
translating them efficiently. Alternatively, the organism can
transcribe a larger number of transcripts and translate each one of
them inefficiently. The former strategy is energetically favorable
since a lower number of transcripts has to be produced; however it was
suggested that it may lead to a noisier pattern of protein expression
\cite{Fraser2004}.

In order to determine the influence of these two strategies on
noise, we ran simulations for different transcription and
translation rates in the base model. For each set of parameters we
computed the coefficient of variation, \toeq~(\ref{eq:CoV}), as
shown in \tofig~\ref{fig:CoV}. As a visual help we also plotted
curves corresponding to equal average levels of $P$, $N$ and $S$.
Examples of stochastic simulations at some extreme values are shown
in the figure, as well.

Two axes of variation can be roughly identified.  Along the main
diagonal (from the bottom left to the upper right corner) the
average level of $N$ grows as the transcription and translation
rates increase. The opposite behavior is seen for $P$, reflecting
the negative feedback loop between both proteins. At the same time,
the coefficient of variation of $P$ remains practically constant
along this axis. Along the other diagonal the coefficient of
variation of $P$ changes from a minimum at high transcription and
low translation rates and reaches a maximum at low transcription and
high translation rates. This behavior is seen for the coefficient of
variation of $N$ as well (data not shown). We also observed that
higher levels of noise are obtained when the average number of mRNA
transcripts is low, and vice versa (see the red equi--$\langle S
\rangle$ lines in \tofig~\ref{fig:CoV}), demonstrating the
importance of the mRNA fluctuations to the excitatory behavior.

Experimental measurements of protein expression reveal that in many
cases the variance in protein levels is roughly proportional to the
mean. This trend is usually explained in terms of an increase in the
transcription noise with the expression level \cite{BarEven2006,
Newman2006}. Yet, some genes are observed to deviate from this
scaling law.

Our results show a general agreement with the previous observation;
at a fixed transcription rate, the noise increases with the
expression level. However, our results also show that noise
fundamentally depends on the interplay between the transcription and
translation rates. For instance, along the lines of constant protein
levels the coefficient of variation reaches a maximum at low
transcription and high translation rates. This suggests that the
variability in protein expression depends not only on the average
expression level, but also on the ratio of transcription versus
translation efficiency. Additional factors such as non--linear
regulatory loops (as in our model system) may also lead to
deviations from the scaling rule.

\subsection{Regulatory RNA filters transcription noise}
\label{results2}

We want to address here the question of how the presence of a
regulatory RNA transcript, which targets the mRNA transcript $S$,
modifies the excitatory behavior inherited from the mRNA stochastic
fluctuations.

\tofig~\ref{fig:examples}(c) shows a typical stochastic simulation
of the extended model, that includes a regulatory RNA, run with the
same biochemical parameters as the base model,
\tofig~\ref{fig:examples}(a).  While pulses of protein expression
were present in both networks, the extended model,
\toeq[s]~(\ref{eq_net_III}), showed broader peaks with significantly
lower amplitude relative to the average expression.  A comparison of
the auto--correlation in both models showed a weaker spiky behavior
in the extended model.  Similarly, while the power spectrum shared a
similar trend with the base model, it lacked the clear maximum
present in the latter one.

The deterministic levels of $P$ (red line) is also shown for both
models in \tofig~\ref{fig:examples}. While none of the systems was
correctly described by the deterministic solution, the departure
from the deterministic dynamics was smaller in the model with
regRNA.
This suggests that, in the presence of a regulatory RNA molecule,
the individual cell's dynamics is more robust to transcript
fluctuations, hence leading to a reduced cell--to--cell variability.

To verify that the reduced excitability in the extended model is due
to the action of the regRNA, we tested the dynamics of the extended
model at different regRNA transcription rates and compared it with
the base model dynamics. If our hypothesis is correct, the two systems
should show differences in their excitatory dynamics as the amount
of regRNA increases.
%
\begin{figure*}
\begin{center}
\includegraphics[width=12cm]{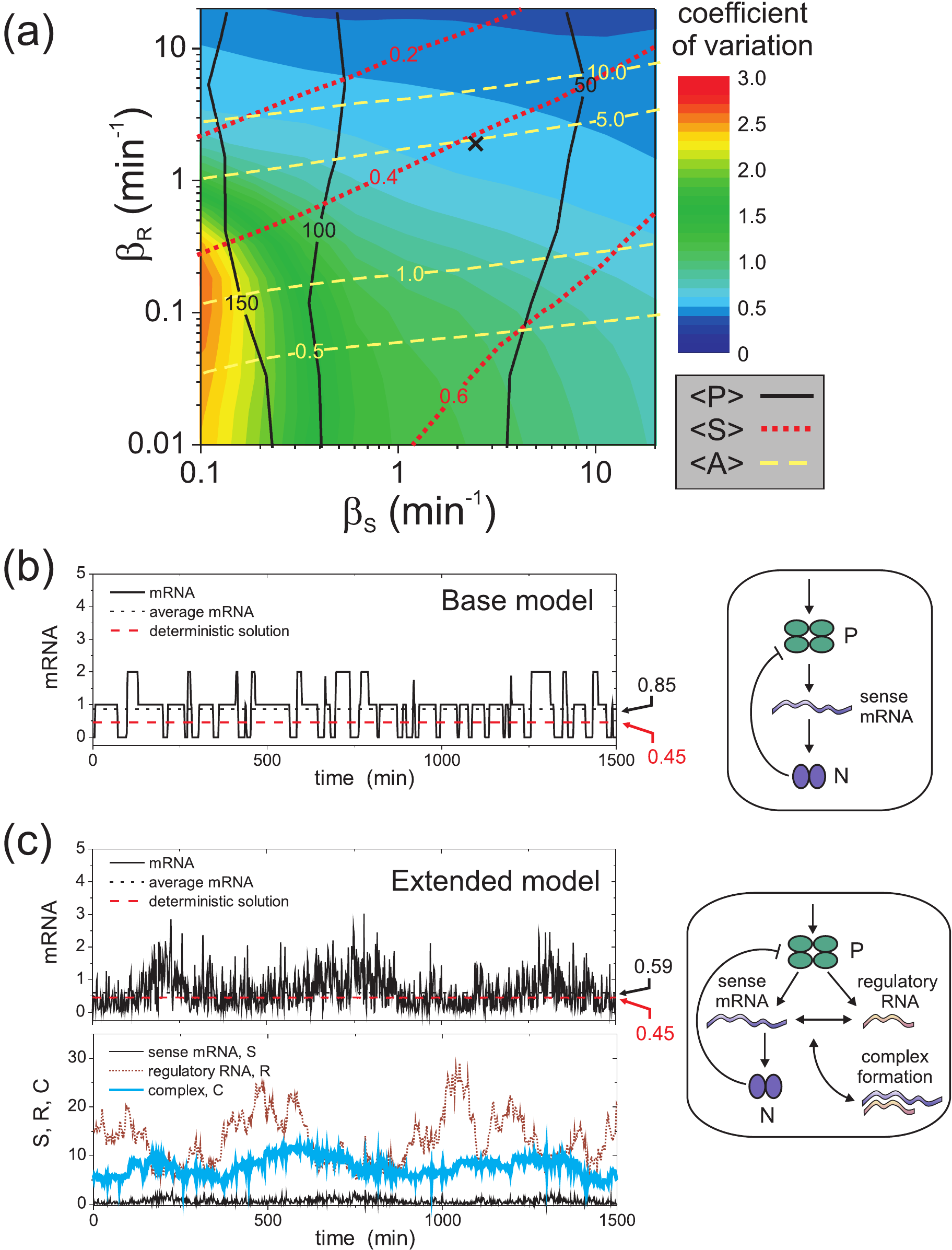}
\caption{\label{fig:buffering} (Color online) Buffering role of the
regulatory RNA in the extended model. (a) Color map of the
coefficient of variation of $P$ in the extended model, as a function
of the sense and regRNA transcription coefficients ($\betaS$ and
$\betaR$ respectivley), with $\lambda=10$.  Contour lines show
equal average levels of $P$ (black), mRNA (dotted red) and regRNA (dashed yellow).
The coefficient of variation is maximized at low transcription
rates.  The black cross indicates the location of the example shown
in \tofig~\ref{fig:examples}(a). (b) Example of the mRNA variation
(black) for a stochastic simulation of the base model with $\betaS =
2.4$ and $\lambda = 5.0$. The dotted black line and the dashed red lines show,
respectively, the average mRNA concentration and the mRNA
deterministic trajectory. (c) Top: mRNA concentration (black) for a
stochastic simulation of the extended model with identical
parameters and $\betaR=2.0$. The mRNA average value is shown with
the dotted black line, and the mRNA deterministic trajectory is
indicated with the dashed red line. Bottom: concentrations of sense mRNA ($S$, thin black),
regRNA ($R$, dotted brown) and $S$--$R$ complex ($C$, thick cyan).  The right
panels show schematic representations of the base and extended
models.}
\end{center}
\end{figure*}
Simulations started at a minimum sense transcription rate of $0.1$
transcripts/min in order to have a sufficient amount of $N$ for
effective repression of $P$. At low regRNA copy numbers, the
coefficient of variation was maximized at low sense transcription
rate, in agreement with the qualitative behavior of the base model.
However, at high regRNA copy number (above $15$ transcripts), the
coefficient of variation became independent of the sense
transcription rate.

The origin of this difference becomes clear when one observes the
time traces of the different RNA transcripts in both models.
\tofig~\ref{fig:buffering}(b) shows the mRNA time trace in the base
model, where no regRNA is present in the system. Clear mRNA micro
pulses with a typical length of the order of the mRNA half--life are
observed. These micro pulses are amplified during translation,
producing the observed pulses of protein expression. An equivalent
plot for the extended model, \tofig~\ref{fig:buffering}(c), reveals
a completely different dynamics. The regRNA serves as a capacitor:
it sequesters sense mRNA when there are copies available, and
releases them back when there are none. Since the typical binding
and unbinding rates are much faster than the RNA half--life, this
sequester and release process is repeated many times during a
typical mRNA micro--pulse, effectively erasing memory from any
previous mRNA stochastic state. For completeness,
\tofig~\ref{fig:buffering}(c) also shows the concentrations of
sense, regRNA and sense--regRNA complex. Notice also that the
average mRNA concentration is significantly closer to the
deterministic trajectory in the extended model, which evidences the
reduction of the stochastic fluctuations in the network.

Based on these observations we conclude that the regRNA is able to filter
out some of the stochastic fluctuations and induce a smoother
protein expression pattern. Such noise dampening capacity was
predicted before in a circuit in which a shared transcription factor
regulates a pair of sense and a highly expressed antisense
transcript \cite{Lapidot2006}.

\begin{figure*}
\begin{center}
\includegraphics[width=11.5cm]{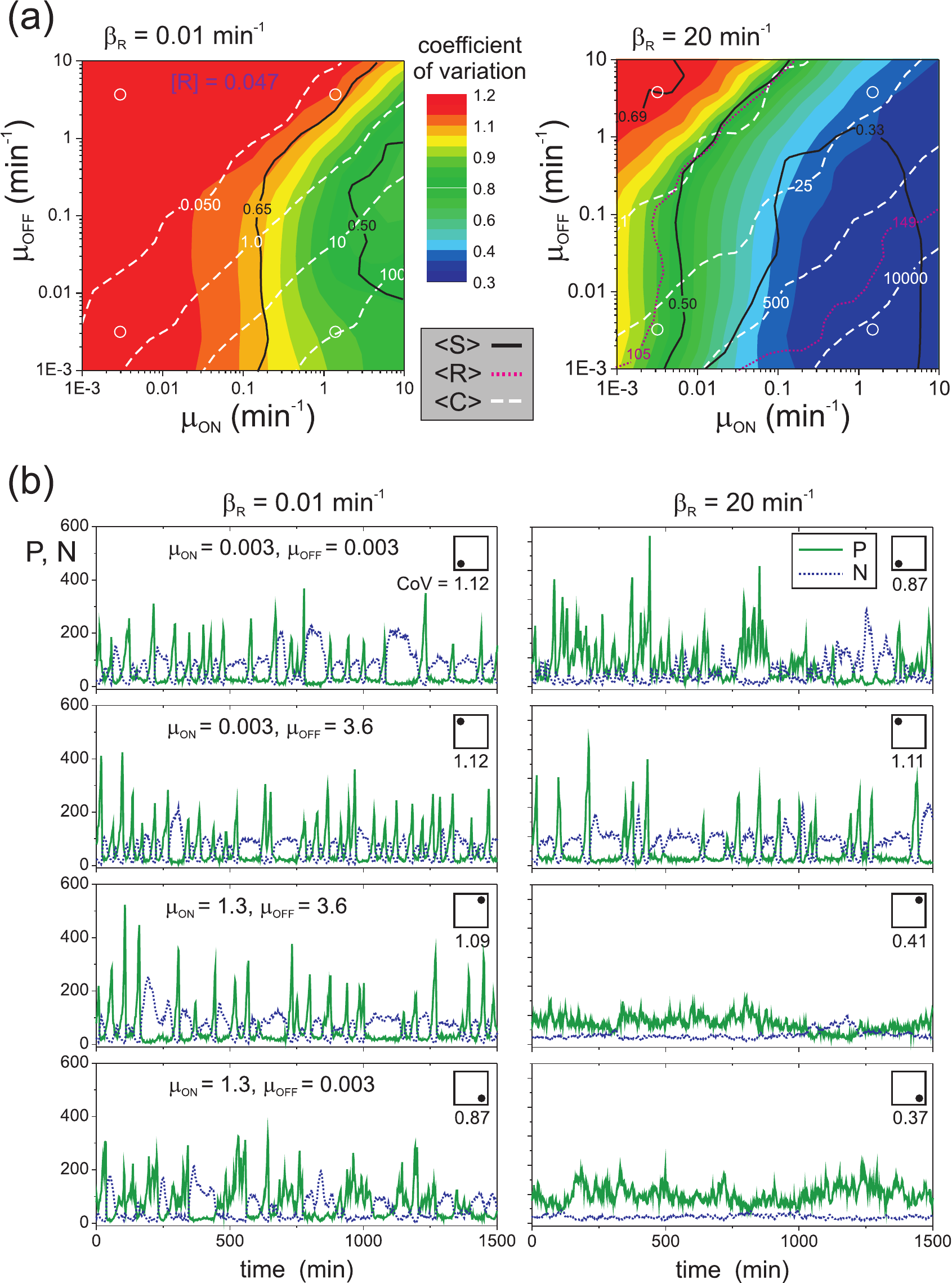}
\caption{\label{fig:complex} (Color online) Effect of the
binding--unbinding rates on the filtering capability of the regRNA.
(a) Color maps of the coefficient of variation of $P$ as a function
 of the binding and unbinding coefficients ($\muON$ and $\muOFF$
respectively) of the sense and regRNA in the extended model. Data
are shown for two different regRNA transcription rates $\betaR$.
Sense transcription and translation rates are maintained constant.
The contour lines show equal average stochastic levels of sense
molecules ($S$, black), regRNA ($R$, dotted pink) and complex ($C$,
dashed white). For $\betaR=0.01$ min$^{-1}$ the number of regRNA is almost
constant and only the average values is shown. (b) Examples of $P$ (thick green)
and $N$ (dotted blue) stochastic simulations for two different values of regRNA
transcription rate $\beta_R$, and for $4$ different combinations of
complex binding ($\muON$) and unbinding rate ($\muOFF$). The sketch
in the top--right of each panel indicates the location of the
examples shown in (a). The value underneath each sketch shows the
corresponding coefficient of variation.}
\end{center}
\end{figure*}

The filtering capability of the regRNA depends crucially on having
fast coupling rates between both RNA strands.
\tofig~\ref{fig:complex}(a) shows the coefficient of variation of
$P$ in the extended model as a function of the sense--regRNA binding
and unbinding rates, and for two different values of the regRNA
transcription rate.

The lowest coefficient of variation is reached in the panel with
higher regRNA transcription rate.  The figure also shows curves of
equal amount of sense $S$, regRNA $R$ and $S$--$R$ complex. In both
panels the minimum coefficient of variation was reached as a
trade--off between having fast binding and unbinding dynamics, and
having the highest number of regRNA molecules (magenta line).  We
conclude that the noise buffering capability needs both a sufficient
number of regRNA transcripts and a fast kinetics to be efficient.
This is clearly illustrated in \tofig~\ref{fig:complex}(b), which
shows examples of stochastic simulations for different combinations
of binding and unbinding rates, and for two extreme values of the
regRNA transcription rate $\betaR$.  Indeed, there is practically no
buffering for low regRNA numbers (left panels), and the time
evolution of both $P$ and $N$ shows a pulsating behavior even at
large binding dynamics. On the contrary, for high regRNA numbers
(right panels), such a pulsating behavior is observed only at very
low binding dynamics, to rapidly disappear as soon as the
binding--unbinding kinetics, and therefore the buffering capacity,
increases.

\section{Discussion}

We have analyzed the response of three different network
architectures of a feedback loop and we have observed remarkable
differences among them that emerge from the RNA dynamical
description.

The fluctuations inherited from the transcription process can be
amplified during translation and produce pulses of protein
expression. The deterministic approach fails to describe such
dynamics both at the single cell level, where no trace of excitatory
behavior is observed, and at the cell distribution level, where the
deterministic steady state underestimates the protein levels.

Deviations between the stochastic average and the deterministic
equations may have different origins \cite{Samoilov2006}. In our
case the difference emerges as a result of the non--linear terms,
which induce correlations that are not present in the deterministic
framework. These correlations become particularly important at low
mRNA levels, where the largest deviation is observed. Indeed,
several works in the literature point at the possibility of having
very few transcripts per cell, in some cases, less than 1 on average
\cite{Holstege1998}. At such low copy numbers, our findings stress
the importance of using stochastic models to accurately describe the
network dynamics.

The excitatory behavior was observed only after simulating the
circuit dynamics with the Gillespie algorithm \cite{Gillespie1977}.
There are alternative methods to model stochastic fluctuations. The
Langevin approach, for instance \cite{Gardiner2004}, adds a small
stochastic term to the continuous deterministic equations to account
 for noise. This approach is not suitable in our system because of
the strong RNA fluctuations, which prevents the characterization of
a smooth continuous background. An alternative approach is based on
the linear noise approximation of the master equation
\cite{VanKampen2007}. It assumes that the system contains a large
number of particles and models noise as a continuous linear gaussian
perturbation. This linearized description has been applied to
processes involving two molecular species \cite{Paulsson04}.
Opposed to the linear approaches, the Gillespie algorithm provides a
simple yet powerful method to obtain stochastic and dynamical
solutions compatible with the full master equation. It does not
require external assumptions on the noise, the number of molecules
and species, or the dynamical regime of the system.

The protein pulses originate in the RNAs stochastic fluctuations.
This result is supported by the observed correlation between the
typical protein pulse length or frequency and the mRNA fluctuations,
as shown in \tofig~\ref{fig:examples}. Additional evidence is provided by
the fact that when the RNA intrinsic randomness is neglected, the
stochastic behavior converges towards the deterministic
concentration. These results are in agreement with experimental
evidences that mRNA fluctuations are a fundamental source of noise
in protein expression \cite{BarEven2006,Newman2006}.

To produce a given average protein concentration the cellular
machinery may choose among different transcription and translation
rates. Minimized variability among cell population is obtained with
a combination of high transcription and low translation rates, as
verified for an auto--regulated gene in steady state using a
linearized stochastic model \cite{Thattai2001}. Our results
generalize this observation. We have analyzed a dynamical,
stochastic negative feedback loop and found, similarly, that noise
is minimized at high transcription and low translation rates, as
shown in \tofig~\ref{fig:CoV}.

The presence of a non--coding regulatory RNA may help buffer the
mRNA fluctuations while allowing the cell to maintain low rates of
transcription. In this work we have considered that a regulatory RNA
(regRNA) binds to the sense transcript and sequesters it from the
cellular environment, thus preventing its translation.  A regRNA
molecule sequesters a mRNA when there are some coding transcripts
available in the medium, and releases them back when there are none.
If this process is repeated sufficient times during a typical mRNA
half--life, then the memory of previous states inherited from the
transcription process can be erased and, therefore, noise can be
partially buffered. In this way, regRNA contributes to reduce the
temporal variability at the single--cell level and, consequently,
also the cell--to--cell variability. A requisite for this mechanism
to work efficiently is fast binding and unbinding (compared to the
typical RNA life--time) between the regulatory RNA and its target
mRNA, as shown in \tofig~\ref{fig:buffering}.
As a result of this buffering, the extended model is less excitable.
The circuit still shows pulses, yet compared to the base model where
there is no regRNA, the pulses are broader and with smaller
amplitude, and they appear at a lower frequency.

A prime example of an oscillating negative feedback loop is provided
by the p53--mdm2 regulatory network. This system has been shown to
oscillate at the single and cell population levels (sustained
oscillations versus damped oscillations respectively)
\cite{Lahav2004, GevaZatorsky2006}.
The maintenance and shape of the oscillations have been linked to
two upstream signaling kinases as well as to an additional negative
feedback loop \cite{Batchelor2008}.
Interestingly however, a single nucleotide polymorphism (SNP309)
found in the mdm2 promoter that results in higher levels of mdm2
mRNA and protein \cite{Bond04} has been shown to disrupt the
oscillations of p53 and mdm2 protein \cite{Hu07}.
This finding supports the idea that RNAs could play an important
role in the oscillatory dynamics in this network, as a high number of
mdm2 mRNAs would minimize the importance of the intrinsic stochastic
fluctuations, and thus attenuate the pulses of protein expression.

While this work has considered a negative feedback loop, the
mechanism that we have described is more general and apply to a wide
variety of regulatory networks. We have shown that the RNA
dynamics is a fundamental source of intrinsic noise, suggesting that
a realistic description of genetic networks requires the stochastic
modeling of the transcription stages of protein expression.

\begin{acknowledgments}
  We thank Arnold Levine, Yael Aylon, Yolanda Espinosa--Parrilla,
  Michal Lapidot and Reut Shalgi for interesting discussions. We
  especially thank Gil Hornung and Orna Dahan for a critical reading
  of the manuscript and for raising many interesting observations.
  We acknowledge the support of the Clore Center for Biological
  Physics, the Center for Complexity Science, the Minerva
  Foundation, the European Research Council (ERC) Ideas Grant and the
  EC FP7 funding (ONCOMIRS, grant agreement number 201102).

\end{acknowledgments}




\begin{thebibliography}{41}
\expandafter\ifx\csname natexlab\endcsname\relax\def\natexlab#1{#1}\fi
\expandafter\ifx\csname bibnamefont\endcsname\relax
  \def\bibnamefont#1{#1}\fi
\expandafter\ifx\csname bibfnamefont\endcsname\relax
  \def\bibfnamefont#1{#1}\fi
\expandafter\ifx\csname citenamefont\endcsname\relax
  \def\citenamefont#1{#1}\fi
\expandafter\ifx\csname url\endcsname\relax
  \def\url#1{\texttt{#1}}\fi
\expandafter\ifx\csname urlprefix\endcsname\relax\def\urlprefix{URL }\fi
\providecommand{\bibinfo}[2]{#2}
\providecommand{\eprint}[2][]{\url{#2}}

\bibitem[{\citenamefont{Elowitz et~al.}(2002)\citenamefont{Elowitz, Levine,
  Siggia, and Swain}}]{Elowitz2002}
\bibinfo{author}{\bibfnamefont{M.}~\bibnamefont{Elowitz}},
  \bibinfo{author}{\bibfnamefont{A.}~\bibnamefont{Levine}},
  \bibinfo{author}{\bibfnamefont{E.}~\bibnamefont{Siggia}}, \bibnamefont{and}
  \bibinfo{author}{\bibfnamefont{P.}~\bibnamefont{Swain}},
  \bibinfo{journal}{Science} \textbf{\bibinfo{volume}{297}},
  \bibinfo{pages}{1183} (\bibinfo{year}{2002}).

\bibitem[{\citenamefont{Bar-Even et~al.}(2006)\citenamefont{Bar-Even, Paulsson,
  Maheshri, Carmi, O'Shea, Pilpel, and Barkai}}]{BarEven2006}
\bibinfo{author}{\bibfnamefont{A.}~\bibnamefont{Bar-Even}},
  \bibinfo{author}{\bibfnamefont{J.}~\bibnamefont{Paulsson}},
  \bibinfo{author}{\bibfnamefont{N.}~\bibnamefont{Maheshri}},
  \bibinfo{author}{\bibfnamefont{M.}~\bibnamefont{Carmi}},
  \bibinfo{author}{\bibfnamefont{E.}~\bibnamefont{O'Shea}},
  \bibinfo{author}{\bibfnamefont{Y.}~\bibnamefont{Pilpel}}, \bibnamefont{and}
  \bibinfo{author}{\bibfnamefont{N.}~\bibnamefont{Barkai}},
  \bibinfo{journal}{Nat.\ Genet.} \textbf{\bibinfo{volume}{38}},
  \bibinfo{pages}{636} (\bibinfo{year}{2006}).

\bibitem[{\citenamefont{Newman et~al.}(2006)\citenamefont{Newman, Ghaemmaghami,
  Ihmels, Breslow, Noble, DeRisi, and Weissman}}]{Newman2006}
\bibinfo{author}{\bibfnamefont{J.}~\bibnamefont{Newman}},
  \bibinfo{author}{\bibfnamefont{S.}~\bibnamefont{Ghaemmaghami}},
  \bibinfo{author}{\bibfnamefont{J.}~\bibnamefont{Ihmels}},
  \bibinfo{author}{\bibfnamefont{D.}~\bibnamefont{Breslow}},
  \bibinfo{author}{\bibfnamefont{M.}~\bibnamefont{Noble}},
  \bibinfo{author}{\bibfnamefont{J.}~\bibnamefont{DeRisi}}, \bibnamefont{and}
  \bibinfo{author}{\bibfnamefont{J.}~\bibnamefont{Weissman}},
  \bibinfo{journal}{Nature} \textbf{\bibinfo{volume}{441}},
  \bibinfo{pages}{840} (\bibinfo{year}{2006}).

\bibitem[{\citenamefont{Raj and van Oudenaarden}(2009)}]{Raj2009}
\bibinfo{author}{\bibfnamefont{A.}~\bibnamefont{Raj}} \bibnamefont{and}
  \bibinfo{author}{\bibfnamefont{A.}~\bibnamefont{van Oudenaarden}},
  \bibinfo{journal}{Annu.\ Rev.\ Biophys.} \textbf{\bibinfo{volume}{38}},
  \bibinfo{pages}{255} (\bibinfo{year}{2009}).

\bibitem[{\citenamefont{Lev Bar-Or et~al.}(2000)\citenamefont{Lev Bar-Or, Maya,
  Segel, Alon, Levine, and Oren}}]{LevBarOr2000}
\bibinfo{author}{\bibfnamefont{R.}~\bibnamefont{Lev Bar-Or}},
  \bibinfo{author}{\bibfnamefont{R.}~\bibnamefont{Maya}},
  \bibinfo{author}{\bibfnamefont{L.}~\bibnamefont{Segel}},
  \bibinfo{author}{\bibfnamefont{U.}~\bibnamefont{Alon}},
  \bibinfo{author}{\bibfnamefont{A.}~\bibnamefont{Levine}}, \bibnamefont{and}
  \bibinfo{author}{\bibfnamefont{M.}~\bibnamefont{Oren}},
  \bibinfo{journal}{PNAS.} \textbf{\bibinfo{volume}{97}},
  \bibinfo{pages}{11250} (\bibinfo{year}{2000}).

\bibitem[{\citenamefont{Tiana et~al.}(2002)\citenamefont{Tiana, Jensen, and
  Sneppen}}]{Tiana2002}
\bibinfo{author}{\bibfnamefont{G.}~\bibnamefont{Tiana}},
  \bibinfo{author}{\bibfnamefont{M.}~\bibnamefont{Jensen}}, \bibnamefont{and}
  \bibinfo{author}{\bibfnamefont{K.}~\bibnamefont{Sneppen}},
  \bibinfo{journal}{Eur.\ Phys.\ J. B} \textbf{\bibinfo{volume}{29}},
  \bibinfo{pages}{135} (\bibinfo{year}{2002}).

\bibitem[{\citenamefont{Nelson et~al.}(2004)\citenamefont{Nelson, Ihekwaba,
  Elliott, Johnson, Gibney, Foreman, Nelson, See, Horton, Spiller
  et~al.}}]{Nelson2004}
\bibinfo{author}{\bibfnamefont{D.}~\bibnamefont{Nelson}},
  \bibinfo{author}{\bibfnamefont{A.}~\bibnamefont{Ihekwaba}},
  \bibinfo{author}{\bibfnamefont{M.}~\bibnamefont{Elliott}},
  \bibinfo{author}{\bibfnamefont{J.}~\bibnamefont{Johnson}},
  \bibinfo{author}{\bibfnamefont{C.}~\bibnamefont{Gibney}},
  \bibinfo{author}{\bibfnamefont{B.}~\bibnamefont{Foreman}},
  \bibinfo{author}{\bibfnamefont{G.}~\bibnamefont{Nelson}},
  \bibinfo{author}{\bibfnamefont{V.}~\bibnamefont{See}},
  \bibinfo{author}{\bibfnamefont{C.}~\bibnamefont{Horton}},
  \bibinfo{author}{\bibfnamefont{D.}~\bibnamefont{Spiller}},
  \bibnamefont{et~al.}, \bibinfo{journal}{Science}
  \textbf{\bibinfo{volume}{306}}, \bibinfo{pages}{704} (\bibinfo{year}{2004}).

\bibitem[{\citenamefont{Lahav et~al.}(2004)\citenamefont{Lahav, Rosenfeld,
  Sigal, Geva-Zatorsky, Levine, Elowitz, and Alon}}]{Lahav2004}
\bibinfo{author}{\bibfnamefont{G.}~\bibnamefont{Lahav}},
  \bibinfo{author}{\bibfnamefont{N.}~\bibnamefont{Rosenfeld}},
  \bibinfo{author}{\bibfnamefont{A.}~\bibnamefont{Sigal}},
  \bibinfo{author}{\bibfnamefont{N.}~\bibnamefont{Geva-Zatorsky}},
  \bibinfo{author}{\bibfnamefont{A.}~\bibnamefont{Levine}},
  \bibinfo{author}{\bibfnamefont{M.}~\bibnamefont{Elowitz}}, \bibnamefont{and}
  \bibinfo{author}{\bibfnamefont{U.}~\bibnamefont{Alon}},
  \bibinfo{journal}{Nat.\ Genet.} \textbf{\bibinfo{volume}{36}},
  \bibinfo{pages}{147} (\bibinfo{year}{2004}).

\bibitem[{\citenamefont{Geva-Zatorsky et~al.}(2006)\citenamefont{Geva-Zatorsky,
  Rosenfeld, Itzkovitz, Milo, Sigal, Dekel, Yarnitzky, Liron, Polak, Lahav
  et~al.}}]{GevaZatorsky2006}
\bibinfo{author}{\bibfnamefont{N.}~\bibnamefont{Geva-Zatorsky}},
  \bibinfo{author}{\bibfnamefont{N.}~\bibnamefont{Rosenfeld}},
  \bibinfo{author}{\bibfnamefont{S.}~\bibnamefont{Itzkovitz}},
  \bibinfo{author}{\bibfnamefont{R.}~\bibnamefont{Milo}},
  \bibinfo{author}{\bibfnamefont{A.}~\bibnamefont{Sigal}},
  \bibinfo{author}{\bibfnamefont{E.}~\bibnamefont{Dekel}},
  \bibinfo{author}{\bibfnamefont{T.}~\bibnamefont{Yarnitzky}},
  \bibinfo{author}{\bibfnamefont{Y.}~\bibnamefont{Liron}},
  \bibinfo{author}{\bibfnamefont{P.}~\bibnamefont{Polak}},
  \bibinfo{author}{\bibfnamefont{G.}~\bibnamefont{Lahav}},
  \bibnamefont{et~al.}, \bibinfo{journal}{Mol.\ Syst.\ Biol.}
  \textbf{\bibinfo{volume}{2}} (\bibinfo{year}{2006}).

\bibitem[{\citenamefont{Batchelor et~al.}(2008)\citenamefont{Batchelor, Mock,
  Bhan, Loewer, and Lahav}}]{Batchelor2008}
\bibinfo{author}{\bibfnamefont{E.}~\bibnamefont{Batchelor}},
  \bibinfo{author}{\bibfnamefont{C.}~\bibnamefont{Mock}},
  \bibinfo{author}{\bibfnamefont{I.}~\bibnamefont{Bhan}},
  \bibinfo{author}{\bibfnamefont{A.}~\bibnamefont{Loewer}}, \bibnamefont{and}
  \bibinfo{author}{\bibfnamefont{G.}~\bibnamefont{Lahav}},
  \bibinfo{journal}{Mol.\ Cell} \textbf{\bibinfo{volume}{30}},
  \bibinfo{pages}{277 } (\bibinfo{year}{2008}).

\bibitem[{\citenamefont{Griffith}(1968)}]{Griffith1968}
\bibinfo{author}{\bibfnamefont{J.}~\bibnamefont{Griffith}},
  \bibinfo{journal}{J. Theor.\ Bio.} \textbf{\bibinfo{volume}{20}},
  \bibinfo{pages}{202 } (\bibinfo{year}{1968}).

\bibitem[{\citenamefont{Elowitz and Leibler}(2000)}]{Elowitz2000}
\bibinfo{author}{\bibfnamefont{M.}~\bibnamefont{Elowitz}} \bibnamefont{and}
  \bibinfo{author}{\bibfnamefont{S.}~\bibnamefont{Leibler}},
  \bibinfo{journal}{Nature} \textbf{\bibinfo{volume}{403}},
  \bibinfo{pages}{335} (\bibinfo{year}{2000}).

\bibitem[{\citenamefont{Ma et~al.}(2005)\citenamefont{Ma, Wagner, Rice, Hu,
  Levine, and Stolovitzky}}]{Ma2005}
\bibinfo{author}{\bibfnamefont{L.}~\bibnamefont{Ma}},
  \bibinfo{author}{\bibfnamefont{J.}~\bibnamefont{Wagner}},
  \bibinfo{author}{\bibfnamefont{J.}~\bibnamefont{Rice}},
  \bibinfo{author}{\bibfnamefont{W.}~\bibnamefont{Hu}},
  \bibinfo{author}{\bibfnamefont{A.}~\bibnamefont{Levine}}, \bibnamefont{and}
  \bibinfo{author}{\bibfnamefont{G.}~\bibnamefont{Stolovitzky}},
  \bibinfo{journal}{PNAS.} \textbf{\bibinfo{volume}{102}},
  \bibinfo{pages}{14266} (\bibinfo{year}{2005}).

\bibitem[{\citenamefont{Vilar et~al.}(2002)\citenamefont{Vilar, Kueh, Barkai,
  and Leibler}}]{Vilar02}
\bibinfo{author}{\bibfnamefont{J.}~\bibnamefont{Vilar}},
  \bibinfo{author}{\bibfnamefont{H.}~\bibnamefont{Kueh}},
  \bibinfo{author}{\bibfnamefont{N.}~\bibnamefont{Barkai}}, \bibnamefont{and}
  \bibinfo{author}{\bibfnamefont{S.}~\bibnamefont{Leibler}},
  \bibinfo{journal}{PNAS.} \textbf{\bibinfo{volume}{99}}, \bibinfo{pages}{5988}
  (\bibinfo{year}{2002}).

\bibitem[{\citenamefont{Ciliberto et~al.}(2005)\citenamefont{Ciliberto, Novak,
  and Tyson}}]{Ciliberto2005}
\bibinfo{author}{\bibfnamefont{A.}~\bibnamefont{Ciliberto}},
  \bibinfo{author}{\bibfnamefont{B.}~\bibnamefont{Novak}}, \bibnamefont{and}
  \bibinfo{author}{\bibfnamefont{J.}~\bibnamefont{Tyson}},
  \bibinfo{journal}{Cell Cycle} \textbf{\bibinfo{volume}{4}},
  \bibinfo{pages}{488} (\bibinfo{year}{2005}).

\bibitem[{\citenamefont{Zhang et~al.}(2007)\citenamefont{Zhang, Brazhnik, and
  Tyson}}]{Zhang2007}
\bibinfo{author}{\bibfnamefont{T.}~\bibnamefont{Zhang}},
  \bibinfo{author}{\bibfnamefont{P.}~\bibnamefont{Brazhnik}}, \bibnamefont{and}
  \bibinfo{author}{\bibfnamefont{J.}~\bibnamefont{Tyson}},
  \bibinfo{journal}{Cell Cycle} \textbf{\bibinfo{volume}{6}},
  \bibinfo{pages}{85} (\bibinfo{year}{2007}).

\bibitem[{\citenamefont{Fran\ifmmode~\mbox{\c{c}}\else \c{c}\fi{}ois and
  Hakim}(2005)}]{Francois2005}
\bibinfo{author}{\bibfnamefont{P.}~\bibnamefont{Fran\ifmmode~\mbox{\c{c}}\else
  \c{c}\fi{}ois}} \bibnamefont{and}
  \bibinfo{author}{\bibfnamefont{V.}~\bibnamefont{Hakim}},
  \bibinfo{journal}{Phys.\ Rev.\ E} \textbf{\bibinfo{volume}{72}},
  \bibinfo{pages}{031908} (\bibinfo{year}{2005}).

\bibitem[{\citenamefont{Thattai and van Oudenaarden}(2001)}]{Thattai2001}
\bibinfo{author}{\bibfnamefont{M.}~\bibnamefont{Thattai}} \bibnamefont{and}
  \bibinfo{author}{\bibfnamefont{A.}~\bibnamefont{van Oudenaarden}},
  \bibinfo{journal}{PNAS.} \textbf{\bibinfo{volume}{98}}, \bibinfo{pages}{8614}
  (\bibinfo{year}{2001}).

\bibitem[{\citenamefont{McKane and Newman}(2005)}]{Mckane2005}
\bibinfo{author}{\bibfnamefont{A.}~\bibnamefont{McKane}} \bibnamefont{and}
  \bibinfo{author}{\bibfnamefont{T.}~\bibnamefont{Newman}},
  \bibinfo{journal}{Phys.\ Rev.\ Lett.} \textbf{\bibinfo{volume}{94}},
  \bibinfo{pages}{218102} (\bibinfo{year}{2005}).

\bibitem[{\citenamefont{McKane et~al.}(2007)\citenamefont{McKane, Nagy, Newman,
  and Stefanini}}]{Mckane2007}
\bibinfo{author}{\bibfnamefont{A.}~\bibnamefont{McKane}},
  \bibinfo{author}{\bibfnamefont{J.}~\bibnamefont{Nagy}},
  \bibinfo{author}{\bibfnamefont{T.}~\bibnamefont{Newman}}, \bibnamefont{and}
  \bibinfo{author}{\bibfnamefont{M.}~\bibnamefont{Stefanini}},
  \bibinfo{journal}{J. Stat.\ Phys.} \textbf{\bibinfo{volume}{128}},
  \bibinfo{pages}{165} (\bibinfo{year}{2007}).

\bibitem[{\citenamefont{Golding et~al.}(2005)\citenamefont{Golding, Paulsson,
  Zawilski, and Cox}}]{Golding2005}
\bibinfo{author}{\bibfnamefont{I.}~\bibnamefont{Golding}},
  \bibinfo{author}{\bibfnamefont{J.}~\bibnamefont{Paulsson}},
  \bibinfo{author}{\bibfnamefont{S.}~\bibnamefont{Zawilski}}, \bibnamefont{and}
  \bibinfo{author}{\bibfnamefont{E.}~\bibnamefont{Cox}},
  \bibinfo{journal}{Cell} \textbf{\bibinfo{volume}{123}}, \bibinfo{pages}{1025
  } (\bibinfo{year}{2005}).

\bibitem[{\citenamefont{Chubb et~al.}(2006)\citenamefont{Chubb, Trcek, Shenoy,
  and Singer}}]{Chubb2006}
\bibinfo{author}{\bibfnamefont{J.}~\bibnamefont{Chubb}},
  \bibinfo{author}{\bibfnamefont{T.}~\bibnamefont{Trcek}},
  \bibinfo{author}{\bibfnamefont{S.}~\bibnamefont{Shenoy}}, \bibnamefont{and}
  \bibinfo{author}{\bibfnamefont{R.}~\bibnamefont{Singer}},
  \bibinfo{journal}{Curr.\ Biol.} \textbf{\bibinfo{volume}{16}},
  \bibinfo{pages}{1018 } (\bibinfo{year}{2006}).

\bibitem[{\citenamefont{Raj et~al.}(2006)\citenamefont{Raj, Peskin, Tranchina,
  Vargas, and Tyagi}}]{Raj2006}
\bibinfo{author}{\bibfnamefont{A.}~\bibnamefont{Raj}},
  \bibinfo{author}{\bibfnamefont{C.}~\bibnamefont{Peskin}},
  \bibinfo{author}{\bibfnamefont{D.}~\bibnamefont{Tranchina}},
  \bibinfo{author}{\bibfnamefont{D.}~\bibnamefont{Vargas}}, \bibnamefont{and}
  \bibinfo{author}{\bibfnamefont{S.}~\bibnamefont{Tyagi}},
  \bibinfo{journal}{PLoS Biol.} \textbf{\bibinfo{volume}{4}},
  \bibinfo{pages}{e309} (\bibinfo{year}{2006}).

\bibitem[{\citenamefont{Bartel}(2004)}]{Bartel2004}
\bibinfo{author}{\bibfnamefont{D.}~\bibnamefont{Bartel}},
  \bibinfo{journal}{Cell} \textbf{\bibinfo{volume}{116}}, \bibinfo{pages}{281}
  (\bibinfo{year}{2004}).

\bibitem[{\citenamefont{Lapidot and Pilpel}(2006)}]{Lapidot2006}
\bibinfo{author}{\bibfnamefont{M.}~\bibnamefont{Lapidot}} \bibnamefont{and}
  \bibinfo{author}{\bibfnamefont{Y.}~\bibnamefont{Pilpel}},
  \bibinfo{journal}{EMBO reports} \textbf{\bibinfo{volume}{7}},
  \bibinfo{pages}{1216} (\bibinfo{year}{2006}).

\bibitem[{\citenamefont{Cawley et~al.}(2004)\citenamefont{Cawley, Bekiranov,
  Ng, Kapranov, Sekinger, Kampa, Piccolboni, Sementchenko, Cheng, Williams
  et~al.}}]{Cawley2004}
\bibinfo{author}{\bibfnamefont{S.}~\bibnamefont{Cawley}},
  \bibinfo{author}{\bibfnamefont{S.}~\bibnamefont{Bekiranov}},
  \bibinfo{author}{\bibfnamefont{H.}~\bibnamefont{Ng}},
  \bibinfo{author}{\bibfnamefont{P.}~\bibnamefont{Kapranov}},
  \bibinfo{author}{\bibfnamefont{E.}~\bibnamefont{Sekinger}},
  \bibinfo{author}{\bibfnamefont{D.}~\bibnamefont{Kampa}},
  \bibinfo{author}{\bibfnamefont{A.}~\bibnamefont{Piccolboni}},
  \bibinfo{author}{\bibfnamefont{V.}~\bibnamefont{Sementchenko}},
  \bibinfo{author}{\bibfnamefont{J.}~\bibnamefont{Cheng}},
  \bibinfo{author}{\bibfnamefont{A.}~\bibnamefont{Williams}},
  \bibnamefont{et~al.}, \bibinfo{journal}{Cell} \textbf{\bibinfo{volume}{116}},
  \bibinfo{pages}{499 } (\bibinfo{year}{2004}).

\bibitem[{\citenamefont{Shalgi et~al.}(2007)\citenamefont{Shalgi, Lieber, Oren,
  and Pilpel}}]{Shalgi2007}
\bibinfo{author}{\bibfnamefont{R.}~\bibnamefont{Shalgi}},
  \bibinfo{author}{\bibfnamefont{D.}~\bibnamefont{Lieber}},
  \bibinfo{author}{\bibfnamefont{M.}~\bibnamefont{Oren}}, \bibnamefont{and}
  \bibinfo{author}{\bibfnamefont{Y.}~\bibnamefont{Pilpel}},
  \bibinfo{journal}{PLoS Comput.\ Biol.} \textbf{\bibinfo{volume}{3}},
  \bibinfo{pages}{e131} (\bibinfo{year}{2007}).

\bibitem[{\citenamefont{Chen et~al.}(2005)\citenamefont{Chen, Sun, Hurst,
  Carmichael, and Rowley}}]{Chen2005}
\bibinfo{author}{\bibfnamefont{J.}~\bibnamefont{Chen}},
  \bibinfo{author}{\bibfnamefont{M.}~\bibnamefont{Sun}},
  \bibinfo{author}{\bibfnamefont{L.}~\bibnamefont{Hurst}},
  \bibinfo{author}{\bibfnamefont{G.}~\bibnamefont{Carmichael}},
  \bibnamefont{and} \bibinfo{author}{\bibfnamefont{J.}~\bibnamefont{Rowley}},
  \bibinfo{journal}{Trends Genet.} \textbf{\bibinfo{volume}{21}},
  \bibinfo{pages}{326} (\bibinfo{year}{2005}).

\bibitem[{\citenamefont{Katayama et~al.}(2005)\citenamefont{Katayama, Tomaru,
  Kasukawa, Waki, Nakanishi, Nakamura, Nishida, Yap, Suzuki, Kawai
  et~al.}}]{Katayama2005}
\bibinfo{author}{\bibfnamefont{S.}~\bibnamefont{Katayama}},
  \bibinfo{author}{\bibfnamefont{Y.}~\bibnamefont{Tomaru}},
  \bibinfo{author}{\bibfnamefont{T.}~\bibnamefont{Kasukawa}},
  \bibinfo{author}{\bibfnamefont{K.}~\bibnamefont{Waki}},
  \bibinfo{author}{\bibfnamefont{M.}~\bibnamefont{Nakanishi}},
  \bibinfo{author}{\bibfnamefont{M.}~\bibnamefont{Nakamura}},
  \bibinfo{author}{\bibfnamefont{H.}~\bibnamefont{Nishida}},
  \bibinfo{author}{\bibfnamefont{C.~C.} \bibnamefont{Yap}},
  \bibinfo{author}{\bibfnamefont{M.}~\bibnamefont{Suzuki}},
  \bibinfo{author}{\bibfnamefont{J.}~\bibnamefont{Kawai}},
  \bibnamefont{et~al.}, \bibinfo{journal}{Science}
  \textbf{\bibinfo{volume}{309}}, \bibinfo{pages}{1564} (\bibinfo{year}{2005}).

\bibitem[{\citenamefont{Gillespie}(1977)}]{Gillespie1977}
\bibinfo{author}{\bibfnamefont{D.}~\bibnamefont{Gillespie}},
  \bibinfo{journal}{J. Phys.\ Chem.} \textbf{\bibinfo{volume}{81}},
  \bibinfo{pages}{2340} (\bibinfo{year}{1977}).

\bibitem[{\citenamefont{Loinger and Biham}(2007)}]{Loinger2007b}
\bibinfo{author}{\bibfnamefont{A.}~\bibnamefont{Loinger}} \bibnamefont{and}
  \bibinfo{author}{\bibfnamefont{O.}~\bibnamefont{Biham}},
  \bibinfo{journal}{Phys.\ Rev.\ E} \textbf{\bibinfo{volume}{76}},
  \bibinfo{pages}{051917} (\bibinfo{year}{2007}).

\bibitem[{\citenamefont{Stommel and Wahl}(2004)}]{Stommel04}
\bibinfo{author}{\bibfnamefont{J.}~\bibnamefont{Stommel}} \bibnamefont{and}
  \bibinfo{author}{\bibfnamefont{G.}~\bibnamefont{Wahl}},
  \bibinfo{journal}{EMBO J.} \textbf{\bibinfo{volume}{23}},
  \bibinfo{pages}{1547} (\bibinfo{year}{2004}).

\bibitem[{\citenamefont{Carter et~al.}(2005)\citenamefont{Carter, Sharov,
  VanBuren, Dudekula, Carmack, Nelson, and Ko}}]{Carter05}
\bibinfo{author}{\bibfnamefont{M.}~\bibnamefont{Carter}},
  \bibinfo{author}{\bibfnamefont{A.}~\bibnamefont{Sharov}},
  \bibinfo{author}{\bibfnamefont{V.}~\bibnamefont{VanBuren}},
  \bibinfo{author}{\bibfnamefont{D.}~\bibnamefont{Dudekula}},
  \bibinfo{author}{\bibfnamefont{C.}~\bibnamefont{Carmack}},
  \bibinfo{author}{\bibfnamefont{C.}~\bibnamefont{Nelson}}, \bibnamefont{and}
  \bibinfo{author}{\bibfnamefont{M.}~\bibnamefont{Ko}},
  \bibinfo{journal}{Genome Biol.} \textbf{\bibinfo{volume}{6}},
  \bibinfo{pages}{R61} (\bibinfo{year}{2005}).

\bibitem[{\citenamefont{Samoilov and Arkin}(2006)}]{Samoilov2006}
\bibinfo{author}{\bibfnamefont{M.}~\bibnamefont{Samoilov}} \bibnamefont{and}
  \bibinfo{author}{\bibfnamefont{A.}~\bibnamefont{Arkin}},
  \bibinfo{journal}{Nat.\ Biotechnol.} \textbf{\bibinfo{volume}{24}},
  \bibinfo{pages}{1235 } (\bibinfo{year}{2006}).

\bibitem[{\citenamefont{Fraser et~al.}(2004)\citenamefont{Fraser, Hirsh,
  Giaever, Kumm, and Eisen}}]{Fraser2004}
\bibinfo{author}{\bibfnamefont{H.}~\bibnamefont{Fraser}},
  \bibinfo{author}{\bibfnamefont{A.}~\bibnamefont{Hirsh}},
  \bibinfo{author}{\bibfnamefont{G.}~\bibnamefont{Giaever}},
  \bibinfo{author}{\bibfnamefont{J.}~\bibnamefont{Kumm}}, \bibnamefont{and}
  \bibinfo{author}{\bibfnamefont{M.}~\bibnamefont{Eisen}},
  \bibinfo{journal}{PLoS Biol.} \textbf{\bibinfo{volume}{2}},
  \bibinfo{pages}{e137} (\bibinfo{year}{2004}).

\bibitem[{\citenamefont{Holstege et~al.}(1998)\citenamefont{Holstege, Jennings,
  Wyrick, Lee, Hengartner, Green, Golub, Lander, and Young}}]{Holstege1998}
\bibinfo{author}{\bibfnamefont{F.}~\bibnamefont{Holstege}},
  \bibinfo{author}{\bibfnamefont{E.}~\bibnamefont{Jennings}},
  \bibinfo{author}{\bibfnamefont{J.}~\bibnamefont{Wyrick}},
  \bibinfo{author}{\bibfnamefont{T.}~\bibnamefont{Lee}},
  \bibinfo{author}{\bibfnamefont{C.}~\bibnamefont{Hengartner}},
  \bibinfo{author}{\bibfnamefont{M.}~\bibnamefont{Green}},
  \bibinfo{author}{\bibfnamefont{T.}~\bibnamefont{Golub}},
  \bibinfo{author}{\bibfnamefont{E.}~\bibnamefont{Lander}}, \bibnamefont{and}
  \bibinfo{author}{\bibfnamefont{R.}~\bibnamefont{Young}},
  \bibinfo{journal}{Cell} \textbf{\bibinfo{volume}{95}}, \bibinfo{pages}{717 }
  (\bibinfo{year}{1998}).

\bibitem[{\citenamefont{Gardiner}(2004)}]{Gardiner2004}
\bibinfo{author}{\bibfnamefont{C.}~\bibnamefont{Gardiner}},
  \emph{\bibinfo{title}{Handbook of stochastic methods for physics, chemistry,
  and the natural sciences}}, Springer series in synergetics
  (\bibinfo{publisher}{Springer-Verlag}, \bibinfo{address}{Berlin New-York},
  \bibinfo{year}{2004}), \bibinfo{edition}{3rd} ed.

\bibitem[{\citenamefont{Van~Kampen}(2007)}]{VanKampen2007}
\bibinfo{author}{\bibfnamefont{N.}~\bibnamefont{Van~Kampen}},
  \emph{\bibinfo{title}{Stochastic Processes in Physics and Chemistry, Third
  Edition (North-Holland Personal Library)}} (\bibinfo{publisher}{{North
  Holland}}, \bibinfo{year}{2007}).

\bibitem[{\citenamefont{Paulsson}(2004)}]{Paulsson04}
\bibinfo{author}{\bibfnamefont{J.}~\bibnamefont{Paulsson}},
  \bibinfo{journal}{Nature} \textbf{\bibinfo{volume}{427}},
  \bibinfo{pages}{415} (\bibinfo{year}{2004}).

\bibitem[{\citenamefont{Bond et~al.}(2004)\citenamefont{Bond, Hu, Bond, Robins,
  Lutzker, Arva, Bargonetti, Bartel, Taubert, Wuerl et~al.}}]{Bond04}
\bibinfo{author}{\bibfnamefont{G.}~\bibnamefont{Bond}},
  \bibinfo{author}{\bibfnamefont{W.}~\bibnamefont{Hu}},
  \bibinfo{author}{\bibfnamefont{E.}~\bibnamefont{Bond}},
  \bibinfo{author}{\bibfnamefont{H.}~\bibnamefont{Robins}},
  \bibinfo{author}{\bibfnamefont{S.}~\bibnamefont{Lutzker}},
  \bibinfo{author}{\bibfnamefont{N.}~\bibnamefont{Arva}},
  \bibinfo{author}{\bibfnamefont{J.}~\bibnamefont{Bargonetti}},
  \bibinfo{author}{\bibfnamefont{F.}~\bibnamefont{Bartel}},
  \bibinfo{author}{\bibfnamefont{H.}~\bibnamefont{Taubert}},
  \bibinfo{author}{\bibfnamefont{P.}~\bibnamefont{Wuerl}},
  \bibnamefont{et~al.}, \bibinfo{journal}{Cell} \textbf{\bibinfo{volume}{119}},
  \bibinfo{pages}{591} (\bibinfo{year}{2004}).

\bibitem[{\citenamefont{Hu et~al.}(2007)\citenamefont{Hu, Feng, Ma, Wagner,
  Rice, Stolovitzky, and Levine}}]{Hu07}
\bibinfo{author}{\bibfnamefont{W.}~\bibnamefont{Hu}},
  \bibinfo{author}{\bibfnamefont{Z.}~\bibnamefont{Feng}},
  \bibinfo{author}{\bibfnamefont{L.}~\bibnamefont{Ma}},
  \bibinfo{author}{\bibfnamefont{J.}~\bibnamefont{Wagner}},
  \bibinfo{author}{\bibfnamefont{J.}~\bibnamefont{Rice}},
  \bibinfo{author}{\bibfnamefont{G.}~\bibnamefont{Stolovitzky}},
  \bibnamefont{and} \bibinfo{author}{\bibfnamefont{A.}~\bibnamefont{Levine}},
  \bibinfo{journal}{Cancer Res.} \textbf{\bibinfo{volume}{67}},
  \bibinfo{pages}{2757} (\bibinfo{year}{2007}).

\end{thebibliography}
\end{document}